\documentclass[aps,twocolumn,showpacs,superscriptaddress,pre,floatfix]{revtex4-2}
\usepackage{amsmath}
\usepackage{amssymb}
\usepackage{dcolumn}%

\usepackage{tikz}
\usepackage{mathtools}
\usepackage{hyperref}
\usepackage{enumitem}
\hypersetup{
    colorlinks,
    linkcolor={blue!80!black},
    citecolor={blue!80!black},
    urlcolor={blue!80!black}
}

\def\rv{J_2}
\def\dlm{\mathrm{d}\hspace{-0.05em}\ln |m|}
\def\dm{\mathrm{d}\hspace{0.05em}|m|}
\def\dU4{\mathrm{d}U_4/\mathrm{d}\beta}
\def\nn{{NN}}
\def\nnn{{NNN}}
\begin{document}

\title{Frustrated Ising model on the honeycomb lattice: Metastability and universality}
\date{\today}

\author{Denis Gessert}
\altaffiliation{Present address: Institut f\"ur Physik, Technische Universit\"at Chemnitz, 09107 Chemnitz, Germany}
\affiliation{Centre for Fluid and Complex Systems, Coventry University, Coventry, CV1 5FB, United Kingdom}
\affiliation{Institut f\"{u}r Theoretische Physik, Universität Leipzig, IPF 231101, 04081 Leipzig, Germany}

\author{Martin Weigel}
\affiliation{Institut f\"ur Physik, Technische Universit\"at Chemnitz, 09107 Chemnitz, Germany}
\affiliation{Physics Department, Emory University, Atlanta, Georgia 30054, USA}

\author{Wolfhard Janke}
\affiliation{Institut f\"{u}r Theoretische Physik, Universität Leipzig, IPF 231101, 04081 Leipzig, 
  Germany}

\begin{abstract}
  We study the Ising model with competing ferromagnetic nearest- and antiferromagnetic next-nearest-neighbor interactions of strengths $J_1 > 0$ and $J_2 < 0$, respectively, on the honeycomb lattice.  For $J_2 > - J_1 / 4$ it has a ferromagnetic ground state, and previous work has shown that at least for $J_2 \gtrsim -0.2 J_1$ the transition is in the Ising universality class. For even lower $J_2$ some indicators pointing towards a first-order transition were reported. By utilizing population annealing Monte Carlo simulations together with a rejection-free and adaptive update, we can equilibrate systems with $J_2$ as low as $-0.23 J_1$. By means of a finite-size scaling analysis we show that the system undergoes a second-order phase transition within the Ising universality class at least down to $J_2 =-0.23 J_1$ and, most likely, for all $J_2 > - J_1 / 4$. As we show here, there exist very long-lived metastable states in this system explaining the first-order like behavior seen in only partially equilibrated systems.        %
\end{abstract}

\maketitle

\section{Introduction}

Frustrated systems~\cite{Diep2012} often exhibit rather rich critical behavior such as reentrant phases~\cite{Azaria1987} or a large ground-state degeneracy~\cite{Espriu2004}, as well as complex dynamical behavior including, e.g., slow relaxation and aging~\cite{Henkel2010}. Frustration itself may be a key ingredient for the occurrence of self-organized complexity in biological systems~\cite{Katsnelson_2018,Wolf2018,Zakany2022}. Thus, understanding the physics of frustrated systems is of fundamental interest. One of the simplest models of such behavior is the classical spin-$1/2$ Ising model with competing ferromagnetic nearest-neighbor (\nn{}) interactions of strength $J_1>0$ and anti-ferromagnetic next-nearest-neighbor (\nnn{}) interactions of strength $J_2<0$. This model has been studied extensively~\cite{Binder1980,Bloete1987,   Malakis2006, Kalz2008, Kalz2011, Kalz2012, Jin2012} for the case of the square lattice, and it is overall well understood, although the debate on some questions remains ongoing~\cite{Hu2021,Li2021a,Yoshiyama2023}. 

More recently, the same model on the honeycomb lattice has attracted some attention~\cite{Bobak2016,Zukovic2020,Zukovic2021,Schmidt2021,Acevedo2021, Corte2021, Gessert2024}. Unlike in nonfrustrated spin systems, the lattice geometry plays a vital role for the level of frustration present and therefore affects the macroscopic behavior of the system. This is illustrated by the textbook example of the antiferromagnetic \nn{} Ising model on square and triangular lattices, respectively. The former maps to the ferromagnetic Ising model by a simple transformation, whereas the latter remains disordered at all temperatures and has a nonzero ground-state entropy per spin~\cite{Wannier1950,*Wannier1973Errata}. In fact, precisely this example illustrates the reason why the behavior of the $J_1$-$J_2$ Ising model is expected to be different on the honeycomb lattice: On the square lattice, the \nnn{} couplings form square lattices, whereas on a honeycomb lattice they form two triangular lattices. Thus, different phenomena may be anticipated for this case, particularly for strong antiferromagnetic \nnn{} interactions. 

The character of the ground states in the honeycomb system depends on the relative strength of $J_1$ and $J_2$~\cite{Katsura1986}. For $J_2 > -|J_1|/4$, depending on the sign of $J_1$, the system either has a ferromagnetically ordered or a N{\'{e}}el-ordered ground state, whereas for $J_2 < -|J_1|/4$ the ground-state manifold is highly degenerate.
As is the case on the square lattice, a simple transformation maps negative $J_1$ to positive ones (and vice versa).
Only quite recently the model was first studied for nonzero temperatures by Bob{\'{a}}k \emph{et al.}~\cite{Bobak2016}. For $J_2 > -|J_1|/4$, using an effective-field theory (EFT) of clusters of different sizes they find evidence of a tricritical point separating a region of second-order transitions for $-0.1 \lesssim J_2/|J_1| \leq 0$ from a regime with first-order transitions for $-0.25 < J_2/|J_1| \lesssim -0.1$.
One of the authors of Ref.~\cite{Bobak2016} subsequently carried out a Monte Carlo (MC) simulation study~\cite{Zukovic2021} in which no tricritical point was observed. Instead it showed that the transition remains of second order and within the Ising universality class at least for $J_2/|J_1| \geq -0.2$. For even lower values of $J_2$ simulations showed hysteresis upon cooling and heating owed to a dramatic increase of the autocorrelation times. This increased range of the second-order regime as compared to that of Ref.~\cite{Bobak2016} is supported by a cluster mean-field study~\cite{Schmidt2021} which finds only second-order signals for the entire range of $-0.25 < J_2/|J_1| \leq 0$.  We recently presented some Monte Carlo results of the scaling of the partition function zeros~\cite{Gessert2024}, which indicated that the system remains in the Ising universality class at least for $J_2/|J_1|\geq -0.22$. For $J_2/|J_1| < - 1/4$, on the other hand, it appears unclear whether there is a phase transition at all. Specific-heat peaks are suggestive of some sort of transition~\cite{Zukovic2022}, and due to the extremely slow relaxation it has been argued that the system shows signs of spin-glass-like ordering~\cite{Zukovic2020}. Results of machine learning studies~\cite{Acevedo2021, Corte2021} were interpreted as suggestive of two distinct phases seen for small system sizes, hence indicative of the existence of a phase transition.

In the present work, however, we focus on the range in $J_2$ with (anti)ferromagnetic order at $T=0$, i.e., $J_2/|J_1| \in (-1/4,0]$.  As one approaches $J_2\searrow -0.25 |J_1|$, standard Monte Carlo techniques such as the Metropolis algorithm are found to be increasingly inefficient due to two effects: First, the introduction of the antiferromagnetic \nnn{} interaction gives rise to local energy minima that simulations tend to get trapped in. Second, as $\rv$ is lowered the transition temperature decreases, which translates into low acceptance rates that are further reducing the efficacy of the Metropolis algorithm.  For the present study we use the population annealing~\cite{Hukushima2003} framework which is well suited for systems with rough free-energy landscapes~\cite{Wang2015} and which has excellent parallel efficiency~\cite{Barash2017}.
To avoid the problem of low acceptance rates, we resort to using the rejection-free $n$-fold way algorithm as local update~\cite{Bortz1975}.
With this computational setup we manage to reliably equilibrate systems with $\rv$ as low as $-0.23|J_1|$ for a range of system sizes which are way out of reach of simple Metropolis simulations. Employing a finite-size scaling (FSS) analysis we determine critical temperatures and exponents, which for all the considered values of $\rv$ are compatible with the Onsager exponents of the regular two-dimensional Ising model.

The rest of the paper is organized as follows. The model, simulation details and measured observables are summarized in Sec.~\ref{sec:modelObservables}. In Sec.~\ref{sec:algo} we introduce the population annealing (PA) algorithm, as well as the adaptations proposed here, and we provide the used simulation parameters. Our results are presented in Sec.~\ref{sec:results}, and a conclusion is provided in Sec.~\ref{sec:conclusion}. 
\section{Model and observables}\label{sec:modelObservables}
\subsection{Model}

\begin{figure}
    \includegraphics[width=0.5\textwidth,trim= 0 2cm 0 2cm, clip]{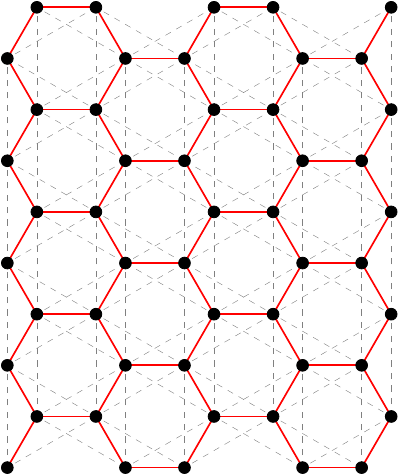}
    \caption{Ising model on a honeycomb lattice. Black circles represent spin sites, solid (respectively dashed) lines nearest-neighbor (respectively next-nearest-neighbor) interactions.}\label{fig:hexagonalLattice}
\end{figure}

In this work we study the two-dimensional $J_1$-$J_2$ Ising model on the honeycomb lattice (see Fig.~\ref{fig:hexagonalLattice}), employing periodic boundary conditions. In the absence of an external magnetic field this model is described by the Hamiltonian
\begin{equation}
    \mathcal{H} = -J_1 \sum_{\langle ij \rangle} \sigma_i \sigma_j - J_2 \sum_{[ ik ]} \sigma_i \sigma_k,\label{eq:Hamiltonian}
\end{equation}
where $\sigma_i\in\{-1,1\}$ are the spin variables. The first sum is over the nearest-neighbor pairs $\langle ij \rangle$ and the second sum relates to next-nearest neighbors $[ik]$. We choose $J_1=1>0$ and $k_B = 1$ to fix units and consider various negative $J_2 \in (-1/4, 0]$. 

The ferromagnetic \nn{} and antiferromagnetic \nnn{} interactions compete with each other, and hence induce frustration, i.e., not all interactions can be satisfied simultaneously. For $J_2 > -1/4$ the ground state consists of all spins aligned with energy per spin $e_\mathrm{FM} = - \frac 3 2 (1 + 2 J_2)$. For stronger antiferromagnetic \nnn{} interactions, i.e., for  $J_2 < -1/4$, the model exhibits a peculiar striped ground state with $e_\mathrm{S} = -\frac 1 2 ( 1 - 2 J_2)$. This state satisfies two of three \nn{} and four of six \nnn{} interactions~\cite{Bobak2016}. In such states certain lines of spins can be flipped at no energetic cost, giving rise to a largely degenerate ground state~\cite{Zukovic2020,Zukovic2022}. Specifically, these states can be mapped to one-dimensional random walks and their degeneracy on a system of edge length $L$ is at least $2^L$~\cite{Gessert2025}.

\subsection{Observables}\label{sec:observables}

To study the critical behavior of the system, we consider various thermodynamic observables. Specifically, from the purely energetic quantities we measure the energy
per spin, $e = \langle\mathcal{H}\rangle / N$, and the specific heat, $C_V = \left(\langle\mathcal{H}^2\rangle - \langle\mathcal{H}\rangle^2\right) / N
T^2$. Here, $T$ denotes the temperature and $N=2L^2$ is the number of spins (see Fig.~\ref{fig:honeycombDomainDecomposition} in Appendix~\ref{app:domainDecompositionHoneycomb}). The purely magnetic quantities considered
are the (absolute) magnetization per spin, $m = \langle |M| \rangle / N$ with $M=\sum_{i=1}^{N} \sigma_i$, the fourth-order Binder cumulant $U_4 = 1 - \langle M^4 \rangle / (3 \langle M^2 \rangle^2 )$, and the magnetic susceptibility. Here, we use both the (connected) magnetic susceptibility $\chi = (\langle M^2\rangle - \langle |M|\rangle^2) / N T$, and the high-temperature (disconnected) magnetic susceptibility $\chi' = \langle M^2\rangle / N T$. Additionally, we consider the mixed quantities,
\begin{equation}
\dm \equiv \frac{1}{N} \frac{\mathrm{d}}{\mathrm{d} \beta} \langle |M| \rangle = \frac{1}{N} \bigl( \langle |M| \rangle \langle \mathcal{H}\rangle - \langle |M| \mathcal{H} \rangle\bigr),
\end{equation}
\begin{equation}
\dlm \equiv \frac{\mathrm{d}}{\mathrm{d} \beta} \ln \langle |M| \rangle = \langle \mathcal{H}\rangle - \frac{\langle |M| \mathcal{H} \rangle}{\langle |M| \rangle},    
\end{equation}
and the derivative of the Binder cumulant
\begin{equation}
    \frac{\mathrm{d} U_4}{\mathrm{d} \beta} = (1-U_4) \left[ \langle \mathcal{H} \rangle - 2 \frac{\langle M^2 \mathcal{H} \rangle}{\langle M^2\rangle} + \frac{\langle M^4 \mathcal{H}\rangle}{\langle M^4\rangle}\right],\label{eq:dU4dbeta}
\end{equation}
with $\beta=1/T$ being the inverse temperature.

For these quantities we expect the following FSS behavior~\cite{Binder1979}:
\begin{subequations}
    \begin{align}
        C_V^{\max} & = c_0 + c_1 \ln L, \label{eq:fss_CV}\\
        \chi_{\max} &  \propto L^{\gamma/\nu}, \label{eq:fss_chi_max}\\
        \chi'(\beta_c) &  \propto L^{\gamma/\nu}, \label{eq:fss_chi_betaC}\\
         \dlm_{\max}& \propto L^{1/\nu}, \label{eq:fss_dlm}\\
         \left[\dU4\right]_{\max} & \propto L^{1/\nu}, \label{eq:fss_dU4db}\\
         \dm_{\max}& \propto L^{(1-\beta)/\nu}, \label{eq:fss_dm}\\
         m(\beta_{\max}^{(\dlm)}) & \propto L^{-\beta/\nu}, \label{eq:fss_m_max}\\
         m(\beta_c) & \propto L^{-\beta/\nu}, \label{eq:fss_m_betaC}
    \end{align}
\end{subequations}
where the subscript ``$\max$'' denotes the peak height of the corresponding quantity %
and $\beta_c$ and $\beta_{\max}^{(\mathcal{O})}$ are the infinite-volume inverse critical temperature and the pseudocritical inverse temperature of the thermodynamic observable~$\mathcal{O}$, respectively. To find $\beta_c$ we use the fitting ansatz~\cite{amit:05}
\begin{equation}
    \beta_{\max}^{(\mathcal{O})}(L) = \beta_c - a L^{-1/\nu} - b L^{-2/\nu} + O(L^{-\omega-1/\nu}) \label{eq:TcFitAnsatz}
\end{equation}
for different observables $\mathcal{O}$. Note that the correction term with the confluent correction exponent $\omega$ is subleading for any $\omega > 1/\nu$. Hence, since $\nu$ is expected to be close to $1$, and $\omega$ to be larger than $1$, we drop this term. Subsequently, we obtain the critical exponents $\beta$, $\gamma$, and $\nu$. Equation~(\ref{eq:fss_CV}) represents a logarithmic divergence of the specific heat in the critical regime.
Note that this corresponds to the case of the two-dimensional Ising universality class, where $\alpha = 0$.

\section{Algorithm and simulation details}\label{sec:algo}
\subsection{Population annealing}\label{sec:algoPA}

Population annealing is an algorithmic framework in which a population of replicas is sequentially cooled while using independent Markov-chain Monte Carlo (MCMC) moves at each temperature, followed by a population control step~\cite{Hukushima2003,machta:10a,Barash2017,Weigel2021}. The algorithm can be summarized as follows:

\begin{enumerate}
\item Initialize a population of $R_0=R$ replicas at the initial inverse temperature   $\beta_0 = 0$. In some cases a nonzero inverse temperature has to be chosen~\cite{Christiansen2019a}. Set iteration counter $i\leftarrow 0$.
\item Repeat the following steps until the final inverse temperature is reached, i.e., until $\beta_i \geq \beta_f$: \begin{enumerate}[label=({\alph*}), ref=2(\alph*)]
  \item Calculate the (normalized) Boltzmann weights for each replica $k$ with energy $E_k$, i.e., $w_{i+1}(E_k) = e^{-\Delta\beta E_k} / \sum_{j=1}^{R_i} e^{-\Delta\beta E_j} $, where $\Delta\beta\coloneqq\beta_{i+1}-\beta_i$ is the inverse temperature step and $R_i$ the population size at $\beta_i$.
  \item Resample the population according to these weights $w_{i+1}(E_k)$, that is, create on average $\tau_{i+1}(E_k) = R_0 w_{i+1}(E_k)$ copies of
    replica~$k$. \label{step:resample}
  \item Increment the iteration counter, $i \leftarrow i+1$.
  \item Perform $\theta$ MCMC sweeps (MCS) (or alternative updates, e.g., for molecular dynamics, see Ref.~\cite{Christiansen2019a}) on each
    replica.\label{step:equilibrate}
  \item Calculate estimates for thermal expectation values of observables $\mathcal O$ through population averages, i.e.,
    \begin{equation}
      \langle \mathcal {O} \rangle \approx \overline{\mathcal{O}} = \frac{1}{R_i} \sum_{k=1}^{R_i} \mathcal{O}_k, \label{eq:popAverage}
    \end{equation}
    where $\langle \dots \rangle$ denotes an expectation value at $\beta_i$. Note that observable $\mathcal{O}$ has to be a configurational quantity~\cite{Ebert2022} that can be calculated for a single configuration such as moments of the energy or magnetization. Estimators for quantities such as $C_V$ and $\chi$ can be constructed from combinations of configurational estimators.
  \end{enumerate}
\end{enumerate}

We choose a population size of $R=20\,000$ throughout. The resampling step~\ref{step:resample} is realized by the nearest-integer resampling method~\cite{Wang2015} given by
\begin{equation}
    P_{\tau_k}(r_k = j) = \begin{cases}
        \tau_k - \lfloor\tau_k \rfloor & \text{if } j = \lfloor\tau_k \rfloor + 1 \\
        1 - (\tau_k - \lfloor\tau_k \rfloor) & \text{if } j = \lfloor\tau_k \rfloor \\
        0 & \text{else}
    \end{cases},
\end{equation}
where the floor function $\lfloor x \rfloor$ denotes the largest integer smaller than or equal to $x$. As we have shown recently~\cite{Gessert2023}, this algorithm is the preferred resampling method for PA when allowing the population size to fluctuate around its target size. As for the annealing schedule, we choose the temperatures adaptively by aiming for constant energy-histogram overlap between two consecutive inverse temperatures $\beta_i$ and $\beta_{i+1}=\beta_i + \Delta\beta$ as suggested by Barash \textit{et al.}~\cite{Barash2017}. The histogram overlap is estimated as
\begin{equation}
  \alpha(\Delta\beta) = \frac{1}{R_i} \sum_{k=1}^{R_i} \min\left(1, \frac{R_0}{R_i} \frac{\exp(-\Delta\beta E_k)}{\sum_{j=1}^{R_i} e^{-\Delta\beta E_j}}\right).
\end{equation}
By means of numerical root finding, we choose $\Delta\beta$ such that $\alpha(\Delta\beta) \approx \alpha^*$ with $\alpha^* = 0.8$. As for the choice of $\theta$, the general advice~\cite{Weigel2021} is to choose it large enough such that a sufficient degree of equilibration is attained. In the past, $\theta$ was usually chosen constant throughout the simulation or temperature-dependent, but fixed prior to the simulation (see for example in Ref.~\cite{Amey2021}). Recently, two of us~\cite{Weigel2021} proposed to also choose $\theta$ adaptively, similarly to how temperature steps are chosen adaptively. We here implement this adaptive equilibration, see Sec.~\ref{sec:adaptiveTheta}.

\subsection{Family quantities, correlation, and population annealing performance}\label{sec:PAQuantities}

Throughout the anneal, the PA performance is monitored by tracking the families of replicas, and related quantities~\cite{Wang2015}. At $\beta_i$, a family refers to a group of replicas that descend from the same replica at $\beta_0$. As replicas may be culled during resampling, the number of nonempty families $f$ is a monotonously decreasing function of $\beta$, and the typical size of such a family increases with~$\beta$. By construction, replicas from different families are uncorrelated. Thus, the family size indicates the maximal correlation within the population. One way to quantify the size of families is via the replica-averaged family size~\cite{Wang2015,Gessert2023}
\begin{equation}
  \rho_t(i) = \frac{1}{R_i} \sum_{k=1}^{R_0} \mathfrak{N}_{i,k}^2,
\end{equation}
where $\mathfrak{N}_{i,k}$ is the number of replicas descending from the initial replica $k$ in the $i$th annealing step at $\beta = \beta_i$. When a PA simulation fails to equilibrate, $\rho_t$ typically increases rapidly. Particularly, when $\rho_t$ reaches the size of the population, that is when all replicas have the same ancestor and there is only one family left, PA results most likely are no longer reliable.

An alternative way to monitor PA performance is through the effective population size $R_\text{eff}(\mathcal{O})$. For a (configurational) observable $\mathcal{O}$~\cite{Ebert2022} the effective population size $R_\text{eff}(\mathcal{O})$ is defined as follows~\cite{Weigel2021},
\begin{equation}
    R_\text{eff}(\mathcal O) = \frac{\sigma^2(\mathcal{O})}{\sigma^2_{R_i}(\overline{\mathcal{O}})},\label{eq:Reff}
\end{equation}
where $\sigma^2(\mathcal{O})$ is the variance of the observable~$\mathcal O$ and $\sigma^2_{R_i}(\overline{\mathcal{O}})$ is the variance of its mean. The first is easily calculated, and the latter can be obtained through binning~\cite{Efron1982}, i.e., 
\begin{equation}
    \hat \sigma^2_{R_i}(\overline{\mathcal O}) = \frac{1}{n (n-1)} \sum_{k=1}^n \left(\mathcal{O}_k^{(n)} - \overline{\mathcal O}\right)^2,\label{eq:varOfMeanBinning}
\end{equation}
where the $n$ blocks are chosen sufficiently large to be effectively uncorrelated and $\mathcal{O}_k^{(n)}$ is the mean of the $k$th block. A decrease in $R_\text{eff}(\mathcal{O})$ may indicate that a PA simulation is falling out of equilibrium. As will be discussed in Sec.~\ref{sec:adaptiveTheta}, we use $R_\text{eff}(E)$ for our adaptive sweep protocol, making sure that this quantity remains almost constant when using this approach.

\subsection{Rejection-free PA}\label{sec:rejectionFreePA}

Next, we replace the most natural choice of the equilibration routine [step~\ref{step:equilibrate} in the PA framework introduced in Sec.~\ref{sec:algoPA}], i.e., the Metropolis algorithm, by the rejection-free $n$-fold way update~\cite{Bortz1975}. For the present model one has $n = (3+1) \times (6+1) \times 2 = 56$ spin classes (3 \nn{}, 6 \nnn{}, and 2 local spin values). In a first attempt one might be tempted to perform $n_\text{f}$ rejection-free updates per site for each replica, implying that each copy ends at a different Metropolis time. As is easily seen, however, this introduces systematic errors particularly at low temperatures~\footnote{For a simple way to see this, consider the   limit of very small, but nonzero temperatures when effectively only the ground   state and the first excited state occur. As there are no available moves with   $\Delta E = 0$, the rejection-free updates lead to an alternation of ground state and first-excited state configurations, even though in Monte Carlo time the state of lower energy is much more likely.}. %
In contrast, when averaging at a fixed Monte Carlo time (as would be the case when using the Metropolis method) the time step $\Delta t$ is accounted for and the obtained estimates are unbiased. In practice, at first $n_\text{f}$ spin flips per site are carried out, such that each replica $k$ arrives at a Metropolis time $t_k$. We then determine $\hat t = \max_k t_k$ and run each replica until it reaches $\hat t$. %
This removes the bias, but comes at the cost of reducing the effective level of parallelism as compared to the Metropolis method, as in the worst case all replicas will need to wait for a single replica to reach $\hat t$.

To summarize, in rejection-free PA one execution of the equilibration routine consists of the steps:
\begin{enumerate}
    \item Perform $n_\text{f}$ $n$-fold way spin updates per site.
    \item Calculate $\hat t = \max_k t_k$.
    \item Perform $n$-fold way spin updates until every replica has reached time $\hat t$.
\end{enumerate}
After completing this update, observables are estimated as population averages as in the original algorithm.
At high temperatures this corresponds to of the order of $n_\text{f}$ MCS whereas at low temperatures it is equivalent to orders of magnitudes more MCS (see Fig.~\ref{fig:performanceSU} in Sec.~\ref{sec:effectOfSweepSchedOnEquil} below).\@

\subsection{Adaptive sweep schedule}\label{sec:adaptiveTheta}

Ideally, the number of performed sweeps should just be large enough to equilibrate and decorrelate the population at each temperature. Too small $\theta$ will cause the simulation to pick up systematic errors and too large $\theta$ will increase the required computational resources without improving the obtained statistics. While it could be somewhat difficult to measure the level of equilibration, we can quite conveniently measure the level of correlation in the population through the effective population size $R_\text{eff}(\mathcal{O})$ defined in Eq.~\eqref{eq:Reff}.

Under the assumption that a high degree of decorrelation is linked to good equilibration, we can define a minimal effective population size $R_\text{eff}^{\min}$ that the population should have before proceeding to the next temperature. In practice, we repeat the equilibration routine until $R_\text{eff}(\mathcal{O})$ is at least $R_\text{eff}^{\min}(\beta_i) = \gamma^{*} R_i$ where $\gamma^*$ (similarly to $\alpha^{*}$) is a constant smaller than $1$, which we here selected to be $0.9$. If the observable $\mathcal{O}$ was chosen to be the magnetization~$M$, then $R_\text{eff}(M)$ would (using Metropolis or $n$-fold way updates) almost surely not reach the target effective population size below the critical temperature due to dynamic ergodicity breaking~\cite{Weigel2021}. We hence choose the energy $E$ as the relevant observable $\mathcal{O}$.

Note that in the above description the number of sweeps necessary is unbounded and thus if the decorrelating effect of the equilibration routine fails to meet the target criterion, the entire simulation will fail to proceed. In a more general setting one should therefore define a cut-off criterion (such as a maximal number of updates or wall-clock time). However, in our case thanks to the combination with rejection-free updates, we did not run into a situation where the adaptive approach spent more time at a temperature than we were willing to allow. Also note that for systems with broken ergodicity this approach is unlikely to be useful and that for such cases an alternative adaptive sweep protocol using the ``restricted autocorrelation time'' has been proposed~\cite{Barzegar2024}.

\subsection{Simulation details}\label{sec:simulationDetail}

\textit{Rejection-free PA} -- We implemented rejection-free PA by means of parallel CPU code using the OpenMP interface running on single compute nodes. Our implementation of the PA framework is based on the open source code provided by the authors of Ref.~\cite{Callaham2017}. As the $n$-fold way update algorithm is inherently difficult to parallelize, we employ one thread per replica. We use $R=20\,000$ as population size, and an adaptive cooling protocol with a target histogram overlap $\alpha^{*} = 0.8$. When the number of sweeps is chosen adaptively, we target an effective population size 
$R_\text{eff}(E)/R_i = \gamma^* = 0.9$. Otherwise, $n_\text{f}$ is chosen constant throughout one simulation and as large as necessary or as large as feasible (whichever is lower). We use the nearest-integer resampling method~\cite{Gessert2023} in step~\ref{step:resample} of the PA algorithm. Error bars are obtained using Eq.~\eqref{eq:varOfMeanBinning}.

\textit{Metropolis PA} -- As a baseline comparison for the proposed adaptive $n$-fold way PA algorithm we employ a graphics processing unit (GPU) implementation of the PA framework based on the code provided in Ref.~\cite{Barash2017}. In addition to the parallelism on the replica-level also present in rejection-free PA, the Metropolis algorithm is parallelized by means of a domain decomposition into 4 sublattices (see Appendix~\ref{app:domainDecompositionHoneycomb}). This highly parallel setting is particularly well-suited for the execution on GPUs. Also here, we use a population size of $R=20\,000$ and the same adaptive temperature protocol. Again, we use the nearest-integer resampling method in step~\ref{step:resample}.

\section{Results}
\label{sec:results}
\subsection{Overview}

\begin{figure}
    \centering
    \includegraphics{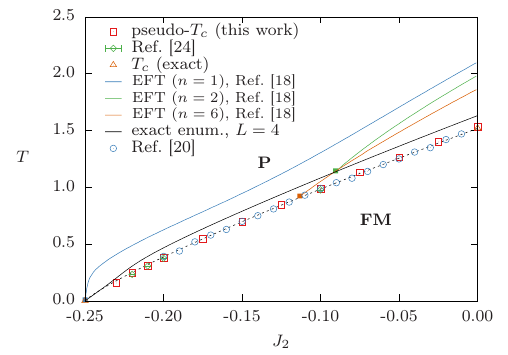}
    \caption{Phase diagram in the $J_2-T$ plane. The square symbols correspond to the pseudocritical temperature obtained from the peak locations of the specific heat for $L=48$ systems. Triangles denote the exactly known values for $\rv = 0$ and $\rv=-1/4$. The solid colored lines show EFT approximations for various cluster sizes $n$ from Ref.~\cite{Bobak2016}, with the end points indicating the predicted tricritical locations. The black solid line shows exact pseudocritical points for an $L=4$ system obtained by exact enumeration. Blue circles show previous MC results from Ref.~\cite{Zukovic2021}. Green diamonds denote FSS extrapolations for $T_c$ from our previous work in Ref.~\cite{Gessert2024} using MC data. The dashed line is merely a guide to the eye. Note that the EFT lines for $n > 1$ do not exist below a certain value of $\rv$, where they (erroneously) predict a tricritical point.}
    \label{fig:critTemp}
\end{figure}

Figure~\ref{fig:critTemp} shows the phase diagram of the system~\eqref{eq:Hamiltonian} with $J_1=1$ in the $\rv - T$ plane for $\rv \geq -1/4$, for which the system undergoes a transition from paramagnetically disordered (P) at high temperatures to ferromagnetically ordered (FM) at low temperatures. In the absence of \nnn{} interactions, i.e., $\rv = 0$, clearly the transition is that of the \nn{} Ising model on the honeycomb lattice in the Ising universality class, and its critical temperature is exactly known, viz.\ $T_c = 2/\ln(2 + \sqrt{3})$~\cite{Houtappel1950}. By adding the antiferromagnetic \nnn{} interaction and increasing its strength, the transition temperature is lowered and vanishes as $\rv \searrow -0.25$. At least for $J_2 \geq -0.2$, previous results~\cite{Zukovic2021} are consistent with the transition remaining within the Ising universality class.

Both our estimates (squares) and those from Ref.~\cite{Zukovic2021} (circles) are obtained through the locations of the specific-heat peaks for $L=48$. For $\rv = -0.1$, $-0.2$, $-0.21$, and $-0.22$ we also show the estimates for $T_c$ of our recent work based on the analysis of partition function zeros as reported in Ref.~\cite{Gessert2024}, which are only slightly lower than these pseudocritical temperatures. Our results of the present work (obtained from rejection-free PA simulations with adaptive sweep schedules) are in very good agreement with previous data. The simulations of Ref.~\cite{Zukovic2021} failed to equilibrate for $\rv < -0.2$ due to very long relaxation times. In Ref.~\cite{Gessert2024}, we employed PA with the same adaptive  sweep protocol as in the present study, however, using the Metropolis update. This approach was already quite costly for $\rv = -0.22$ and failed for even lower values of $\rv$. Here we can equilibrate $\rv \geq -0.23$ thanks to the rejection-free Monte Carlo updates~\cite{Bortz1975} (see Sec.~\ref{sec:rejectionFreePA}). By means of FSS (see Sec.~\ref{sec:fss}) we show that also for $-0.23 \leq \rv < -0.2$ the critical exponents are consistent with the Ising ones.

\begin{figure*}[ht]
  \includegraphics{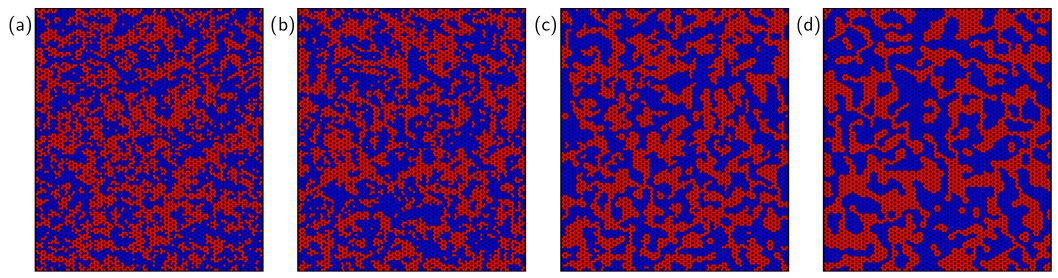}
  \caption{Equilibrium snapshots at the inverse temperature $\beta=0.6 \beta_c$ for $L=64$, and (a) $\rv=0$, (b) $\rv=-0.1$, (c) $\rv=-0.2$, and (d) $\rv=-0.23$. In (c) and (d), for $\rv$ close to $-1/4$, the structures appear much more stable, consistent with the observed metastability and slow relaxation at these values of $\rv$.\label{fig:configs_0.6betaC}}
\end{figure*}

For comparison, the estimates of the critical temperature from an EFT study~\cite{Bobak2016} with varying cluster sizes $n$ are also included. Only when single-spin clusters are used, the method predicts a second-order transition on the full range of $\rv > -0.25$, and otherwise lines end in a tricritical point (indicated by a point), which appears to move towards lower $J_2$ as $n$ increases. Thus, using larger clusters might yield a picture consistent with a much larger region with a second-order transition~\cite{Zukovic2021,Schmidt2021,Gessert2024}. In fact, this behavior is somewhat different from the analogous model on the square lattice: there, even for single-spin clusters an EFT approximation predicts a tricritical point on either side of the special point $\rv^{*}=-0.5$~\cite{Abalmasov2023a,Note10}\footnotetext[10]{Note that in Ref.\ \cite{Abalmasov2023a} the random local field approximation is used, which for uniform polarization is equivalent to single-cluster EFT, as one can show.} which, however, also comes closer to $\rv^{*}$ as the cluster size is increased~\cite{Jin2013,Bobak2015}.

For even smaller values of $\rv$, also our rejection-free approach fails. Therefore, exact enumeration for a very small system of $4 \times 4$ hexagons (corresponding to a 32 spin system) was carried out, and the specific-heat peak locations were determined (black solid line). For this small system the peak location smoothly approaches zero temperature as $\rv \searrow -0.25$, which may indicate that this is also the case in the thermodynamic limit. %

One reason for the slow relaxation of the Metropolis dynamics for lower values of $\rv$ is that the transition takes place at a low temperature: Assuming typical $\Delta E$'s to be of the same order as for larger $\rv$, this leads to small acceptance rates, and therefore unavoidably to slower dynamics. This effect is reduced by using the rejection-free update (see Sec.~\ref{sec:rejectionFreePA}). A slightly more subtle effect also contributing to this slow relaxation is that paramagnetic structures formed at the same value of $\beta/\beta_c$ are more stable for lower values of $\rv$; see Fig.~\ref{fig:configs_0.6betaC} showing configuration snapshots for $\rv=0$,$-0.1$,$-0.2$, and $-0.23$ for $\beta = 0.6\beta_c$. While the configurations in Figs.~\ref{fig:configs_0.6betaC}(a) and \ref{fig:configs_0.6betaC}(b) are very noisy, Figs.~\ref{fig:configs_0.6betaC}(c) and \ref{fig:configs_0.6betaC}(d) show a clear domain picture with smooth boundaries. Using single spin-flip dynamics, overturning such domains is only possible by passing through configurations of much higher energy, rendering these states rather stable (albeit only metastable below the critical temperature). The reason why also the rejection-free approach fails to overturn these domains for low enough values of $\rv$, is that it ``flickers'' between various excited states, that is, one effectively stays in the same subset of spin configurations for a long time. This is illustrated using the example of a single overturned hexagon later in Sec.~\ref{sec:energyBarriers}. 
    
\subsection{Effect of the update and sweep schedule on equilibration}\label{sec:effectOfSweepSchedOnEquil}

\begin{figure}
    \centering
    \includegraphics{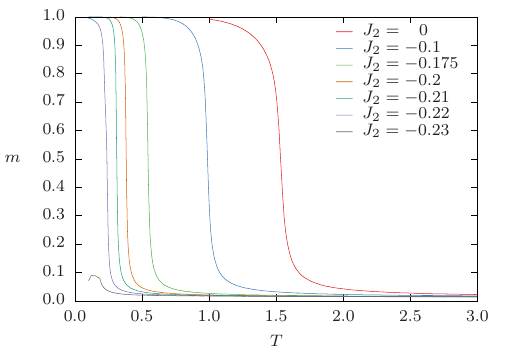}
    \includegraphics{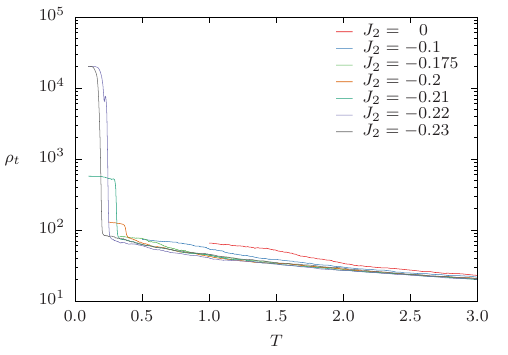}
    \caption{Results from PA simulations using Metropolis updates with $\theta =
      2000$ (respectively $\theta=500$ for $\rv=0$) and $L=48$. Top panel: Magnetization per spin $m$ for various coupling strengths. For $\rv \gtrsim -0.2$ we obtain good agreement with the rejection-free data (not shown here). For $\rv < -0.22$ the simulations fail to reach the fully ordered state at low temperatures. Bottom panel: Replica-averaged family size $\rho_t$. As expected~\cite{Wang2015} $\rho_t$ monotonously increases during the annealing process with decreasing temperature. The failure to equilibrate for $\rv < -0.2$ is reflected in a sharp increase in $\rho_t$ near the corresponding critical temperatures $T_c$ for $\rv \leq -0.2$.}\label{fig:resMetropolis}
\end{figure}

In the following, we present data of PA simulations for systems with $L=48$ using i) the Metropolis algorithm with a fixed number of sweeps, ii) the $n$-fold way update using a fixed number of spin flips before time synchronization, and iii) the $n$-fold way update with an adaptive number of sweeps. In the first case we  encountered severe difficulties in equilibrating simulations. This problem is significantly reduced by using the rejection-free update instead. Finally, the level of equilibration is further improved by deploying the adaptive sweep protocol.

Starting with data from PA simulations using the Metropolis update, Fig.~\ref{fig:resMetropolis} shows the magnetization and replica-averaged family size obtained from PA simulations with $\theta = 2000$ sweeps per replica at each temperature for all $\rv$ (except for $\rv=0$, where $\theta = 500$ is quite sufficient). As physically expected, the magnetization per spin (shown in the top panel) is close to one (zero) below (above) the transition temperature, which decreases as $\rv\searrow -0.25$. This is seen for all shown $\rv$ with the exception of $\rv=-0.23$. In this case the simulation does not reach a fully magnetized state at low temperatures, which indicates that it failed to equilibrate there.

A useful way to detect a failed PA simulation is by considering the replica-averaged family size $\rho_t$. Here, for $\rv=-0.23$ the value of $\rho_t$ at the lowest temperature is equal to the population size $R=20\,000$, which is indicative of a PA simulation which failed to equilibrate. The same is observed for $\rv=-0.22$, despite still reaching the ferromagnetic ground state (hence this is a marginal case). 
A closer look (and in particular a comparison with the better equilibrated $n$-fold way data in Fig.~\ref{fig:MECCoverview} below), however, indeed reveals that for magnetizations larger than $m \approx 0.5-0.6$ the Metropolis data for $\rv=-0.22$ are not reliable.
Also for $\rv\in \{ -0.2, -0.21\}$ a sharp increase in $\rho_t$ is seen near criticality. In these cases, however, the final value of $\rho_t$ is still well below the population size. For the remaining $\rv$, $\rho_t$ monotonously increases with decreasing temperature and no sharp increase at $T_c$ is visible, as would be expected when $\theta$ is larger than the autocorrelation time~\cite{Gessert2023}. %

\begin{figure}
    \centering
    \includegraphics{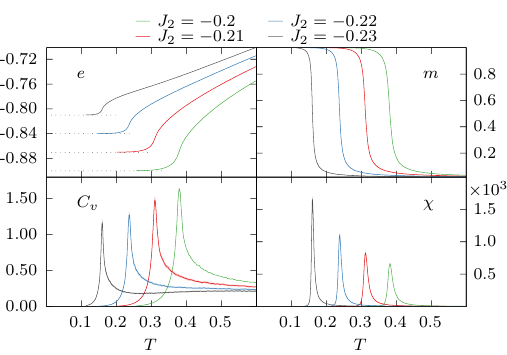}
    \caption{Overview over the thermodynamic quantities measured as functions of temperature for $\rv \in \{-0.2, -0.21, -0.22, -0.23\}$ for a system of linear size $L=48$ from PA simulations using $n$-fold way updates.
    The shaded areas indicate $2\sigma$ error environments of the data for $\rv \in \{-0.21, -0.22, -0.23\}$ (but note that in many cases they are hardly visible). Top left: Energy per spin (dashed lines correspond to the ground-state energy), Top right: Magnetization per spin. Bottom left: Specific heat. Bottom right: Magnetic susceptibility.}\label{fig:MECCoverview}
\end{figure}
In contrast, when using rejection-free PA, the ferromagnetic ground state is reached for all $\rv$ considered above (see Fig.~\ref{fig:MECCoverview}). In this case, we do not use the adaptive sweep schedule yet, but fix $n_\text{f}=500$, that is for each replica 500 spin flips per site were carried out before running further updates to synchronize the Metropolis times of all replicas (see Sec.~\ref{sec:rejectionFreePA}). Jumps in the energy curve~\cite{Zukovic2021}, or kinks in the specific heat and magnetic susceptibility which we observed in less well equilibrated simulations are absent here.

\begin{figure}
    \includegraphics{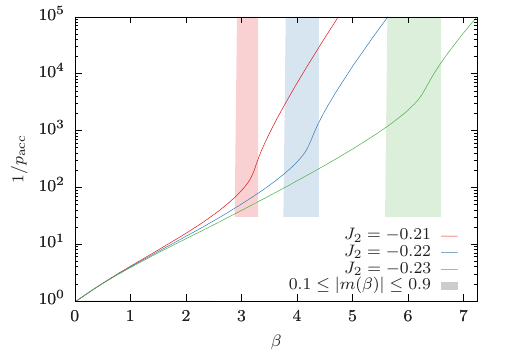}
    \includegraphics{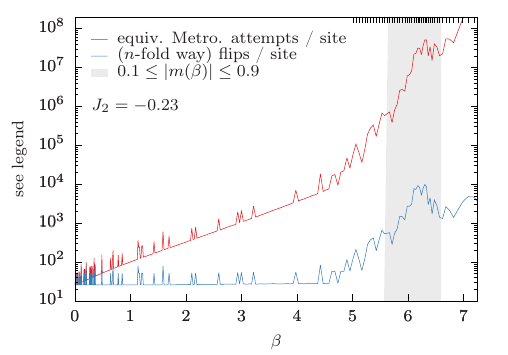}
    \caption{Demonstration of efficacy of rejection-free PA with an adaptive number of sweeps at each temperature using $L=48$. Top panel: Inverse of the Metropolis acceptance probability (which is proportional to the speedup achieved by the $n$-fold way update) as a function of inverse temperature. $1/p_\text{acc}$ grows rapidly in $\beta$ throughout with a marked increase around criticality (colored, shaded regions). Bottom panel: Equivalent number of Metropolis attempts and number of flips made at each inverse temperature using the adaptive sweep protocol for $\rv=-0.23$. For $\beta > 5$ one notes a significant increase in the number of flips. In this range, the inverse temperatures used in the PA run are indicated as $x$ ticks at the top of the plot.}\label{fig:performanceSU}
\end{figure}

The reason why the rejection-free approach improves equilibration is readily understood by considering the Metropolis acceptance rates $p_\text{acc}$ as a function of inverse temperature $\beta$, which across the critical regime decrease at least exponentially in $\beta$ before reaching the asymptotic behavior $p_\text{acc}=\exp(-(6+12 \rv)\beta)$ for large $\beta$ (top panel in Fig.~\ref{fig:performanceSU}). The colored, shaded areas indicate the critical regimes for the different $\rv$. While for $\rv=-0.21$ still 1 in 100 spin flip attempts is accepted, it is only 1 in 5000 for $\rv=-0.23$, rendering the 2000 sweeps per temperature from the Metropolis PA simulations discussed above much too few to decorrelate spin configurations between temperature steps. In fact, $1/p_\text{acc}$ is directly proportional to the computational speedup achieved by using the rejection-free approach as opposed to normal Metropolis. Note that the shown acceptance rates were obtained already using the adaptive sweep schedule (see Sec.~\ref{sec:adaptiveTheta}). (Since the acceptance probability is a thermodynamic observable, it is agnostic to the choice of the sweep schedule.)

\begin{figure}
    \includegraphics{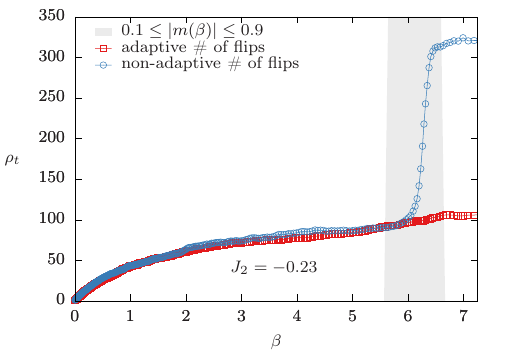}
    \caption{Family growth quantified by $\rho_t$ using respectively not using the adaptive sweep protocol in rejection-free PA. For small $\beta$ both protocols result in similar values for $\rho_t$. When \emph{not} using the adaptive protocol, around criticality $\rho_t$ exhibits a severe jump which is absent when applying the adaptive protocol.}\label{fig:pa-nf-rhoT-adaptvsnonaSw}
\end{figure}

The bottom panel of Fig.~\ref{fig:performanceSU} shows for $\rv = -0.23$ the number of $n$-fold way flips carried out per site and temperature, as well as the equivalent number of spin-flip attempts in the Metropolis scheme using the same simulation data as in the top panel. Specifically, the adaptive sweep criterion of $R_\text{eff}(E) > 0.9 R_i$ was imposed. As one can see, for small $\beta$ only very few flips are required to decorrelate the replicas (25 was chosen as a lower bound), whereas for $\beta \approx 5$ and higher many more flips are required. The adaptive sweep criterion has two benefits: On the one hand, the simulation quickly passes through the easily-to-equilibrate high temperatures, and on the other hand, most time is spent in the much harder-to-equilibrate critical regime and at lower temperatures.  Note also that the adaptively chosen temperatures (using $\alpha^*=0.8$) are denser in the critical region (indicated by the $x$ ticks at the top of the plot for $\beta>5$). Within this region, in total $5\times 10^{8}$ Metropolis MCS are carried out per replica, which corresponds to only $10^5$ rejection-free spin flips per site and replica (making up 55\% of the overall number of spin flips of this PA run).  One can use adaptive sweeps also with regular Metropolis, which in fact is more efficient than without an adaptive schedule. However, with such low acceptance rates it takes a very long time to equilibrate at low temperatures (see equivalent Metropolis attempts in Fig.~\ref{fig:performanceSU}). As mentioned above, our previous simulations of Ref.~\cite{Gessert2024} were carried out using this approach, and equilibrating systems with $\rv=-0.22$ was difficult but still possible (also note the smaller system sizes there). For $\rv=-0.23$, on the other hand, the adaptive sweep protocol using the Metropolis update fails to reach temperatures below the critical one within a reasonable time even for small systems. Note that the visible jumps in the lower panel are artifacts of the discretization used in the sweep protocol.

Last, the replica-averaged family size $\rho_t$ (Fig.~\ref{fig:pa-nf-rhoT-adaptvsnonaSw}) is studied for two rejection-free runs, one with adaptive sweep schedule and one without. First, the adaptive one was run, and then the average number of flips per temperature was calculated, to ascertain that both runs use approximately the same CPU time. For small $\beta$ both curves have similar values, indicating that at high temperature indeed very few spin flips are required for equilibration. Close to criticality, $\rho_t$ of the run without an adaptive sweep schedule then increases rapidly, whereas it continues without a steep increase otherwise. Thus, at the same computational cost much better equilibration is to be expected from the adaptive run.

\subsection{Finite-size scaling analysis}
\label{sec:fss}

We now turn to studying the critical behavior of the ferromagnetic ordering by means of an FSS analysis~\cite{Binder1979} for $\rv=-0.21$, $-0.22$, and $-0.23$. All simulations for $L=12$, $16$, $24$, $32$, $48$, $64$, $88$, and $128$ were carried out using rejection-free PA with the adaptive sweep schedule.

\begin{figure*}[ht]
    \centering
    \includegraphics{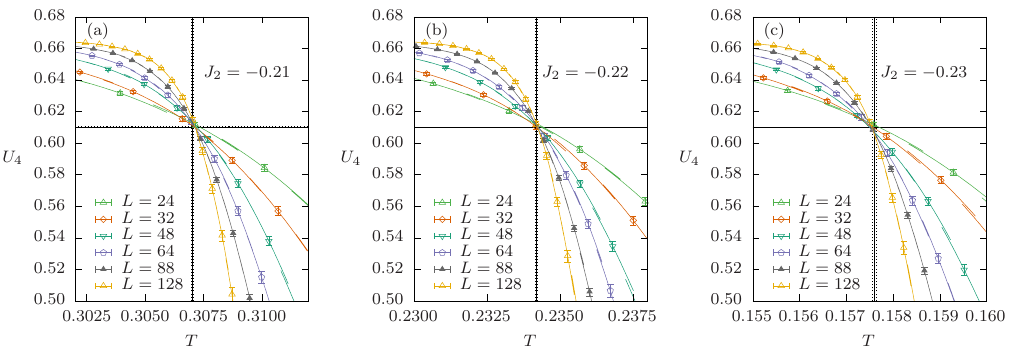}
    \caption{The Binder parameter $U_4$ in the vicinity of $T_c$ for different system sizes and for (a) $J_2=-0.21$, (b) $-0.22$, and (c) $-0.23$, respectively. Points denote the simulation temperatures and the connecting lines were calculated using single-histogram reweighting. The estimated critical temperatures $T_c(\rv)$ are indicated by solid vertical lines and their margins of error by dashed vertical lines. The values for $1/T_c$ of $3.2572(5)$ and $4.2702(6)$ for $J_2=-0.21$ and $-0.22$, respectively, are taken from Ref.~\cite{Gessert2024}, while the value of $6.3455(14)$ for $J_2=-0.23$ is from Table~\ref{tab:fssExponents}. Similarly, the estimated values of $U^*=U_4(T_c)$ (including their error margins) are shown by horizontal lines.\label{fig:BinderParam}}
\end{figure*}
We begin the analysis by considering the Binder parameter $U_4$ and its crossings for different system sizes $L$ at $(T_c,U^*)$ for $\rv=-0.21$, $-0.22$, and $-0.23$  shown in Fig.~\ref{fig:BinderParam}. The data points with error bars correspond to the estimates at each annealing step, while the lines are extrapolations using (single) histogram reweighting~\cite{Ferrenberg1988,*Ferrenberg1989}. Since the individual simulation points are almost fully uncorrelated, the lines do not meet perfectly but instead small jumps of the order of the error bars are visible. As expected, the temperature curves of $U_4$ for different system sizes calculated in this way approximately cross at a single point $(T_c,U^*)$, providing a convenient and well-accessible estimate of the critical temperature $T_c$. Due to finite-size corrections, this crossing is not perfect and we have omitted the data for the two smallest system sizes considered, $L=12$ and $L=16$, as the crossings there were significantly shifted with respect to $T_c$. The vertical lines indicate the previously obtained critical temperature (with dashed lines as error margins), which for $\rv=-0.21$ and $-0.22$ are the values of Ref.~\cite{Gessert2024}, and for $\rv=-0.23$ is given by the value from Table~\ref{tab:fssExponents} from the peak locations of $\dm$ discussed below. In all three cases, the crossings are in good visual agreement with these values for~$T_c$. For $U^*=\lim_{L\rightarrow\infty}U_4(T_c,L)$ and for $\rv=-0.21$, $-0.22$, and $-0.23$, we obtain the values $0.6124(12)$, $0.6120(11)$, and $0.6117(11)$, respectively; see Appendix~\ref{app:binderCrossing} for details.

\begin{figure}
    \centering
    \includegraphics{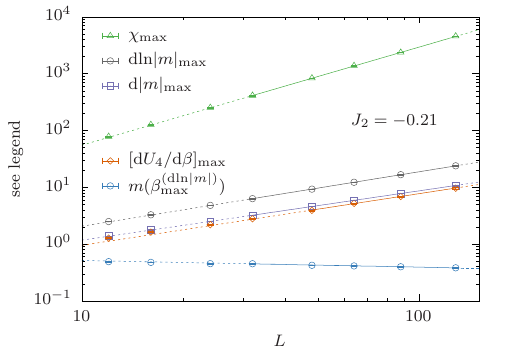}
    \includegraphics{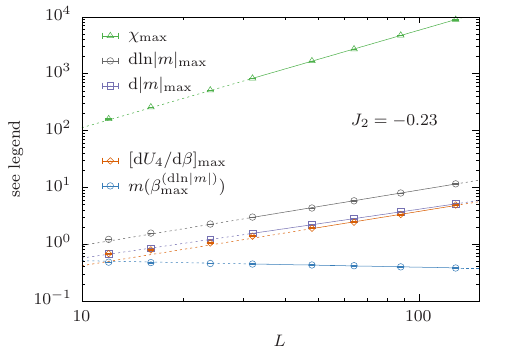}
    \caption{Finite-size scaling fits using Eq.~\eqref{eq:fss_chi_max} and Eqs.~\eqref{eq:fss_dlm}-\eqref{eq:fss_m_max} for $\rv=-0.21$ and $-0.23$ for system sizes up to $L=128$. Solid lines correspond to the fit ranges used to obtain the exponents compiled in Table~\ref{tab:fssExponents}. Dotted lines are extrapolations of the fits to values of $L$ outside the fitting range.}
    \label{fig:fssFits}
\end{figure}

Figure~\ref{fig:fssFits} shows for $\rv=-0.21$ and $\rv=-0.23$ the peak heights of $\chi$, $\dlm$, $\dm$, and $\dU4$, as well as $m$ evaluated at the inverse temperature of the $\dlm$ peak; the values were determined through (single) histogram reweighting~\cite{Ferrenberg1988,*Ferrenberg1989} of the PA data~\footnote{Note that the simulation temperature closest to the corresponding peak location is used as reference temperature in histogram reweighting. Alternatively, it would also be possible to use the populations from all temperatures in a multi-histogram approach, see Ref.~\cite{Barash2017}.}. The finite-size scaling relations~\eqref{eq:fss_CV}-\eqref{eq:fss_m_betaC} only hold asymptotically, and in principle one expects (unknown) correction terms that vanish as $L\rightarrow\infty$.
To reduce the systematic error from omitting these finite-size corrections, system sizes $L<32$ were discarded in the fits, such that all fits have a quality of fit $Q > 0.1$ -- except for $\dU4$ for which also $L=32$ had to be discarded in order to satisfy $Q>0.1$. While we will elaborate on the fitting results in more detail below in the discussion of Table~\ref{tab:fssExponents}, note that the obtained values for $1/\nu$ using Eq.~\eqref{eq:fss_dlm} of 0.96(1) for $\rv=-0.21$, and 0.98(1) for $\rv=-0.23$ are consistent with the expected Onsager value of 1, albeit not fully within error bars.

\begin{figure}
    \centering
    \includegraphics{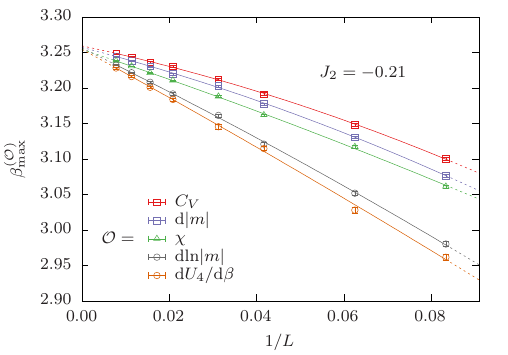}
    \includegraphics{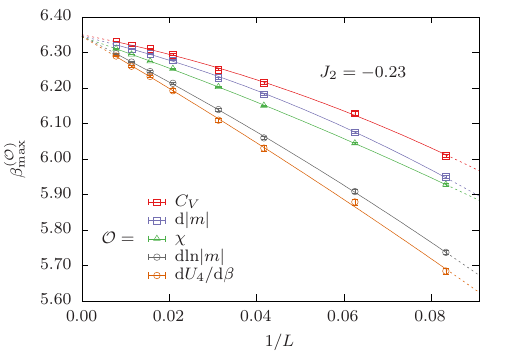}
    \caption{Finite-size scaling fits of the peak locations for $\rv=-0.21$ and $-0.23$ for system sizes up to $L=128$. Solid lines correspond to the fit ranges used to obtain the $\beta_c$ estimates quoted in Table~\ref{tab:fssExponents}.
    }
    \label{fig:fssFitsTc}
\end{figure}

To estimate the inverse critical temperature $\beta_c(\rv)$, the ansatz~\eqref{eq:TcFitAnsatz} is used for each observable independently on the full range of system sizes, i.e., on $L \in [12,128]$. Figure~\ref{fig:fssFitsTc} shows the location of the pseudocritical points, as well as the fits of Eq.~\eqref{eq:TcFitAnsatz} for $\rv=-0.21$ and $\rv=-0.23$, and fixing $\nu=1$. (We fixed $\nu=1$ since the numerical data of the values of $\dlm$ at the maxima are consistent with that value, and since it also is the expected value of the Ising universality class.) In both cases the obtained values of $\beta_c$ are compatible within error bars for different observables (see Table~\ref{tab:fssExponents}). From the visible curvature of the curves in Fig.~\ref{fig:fssFitsTc} one notes that clearly a correction term such as the one in Eq.~\eqref{eq:TcFitAnsatz} is necessary to describe $\beta^{(\mathcal{O})}_{\max}(L)$ --- the exception being $\dlm$ whose pseudocritical points almost fall on a straight line.

\begingroup
\squeezetable
\begin{table*}[ht]
    \caption{Critical exponents and corresponding estimates for the inverse critical temperature $\beta_c$ determined through FSS fits for $\rv=-0.21$, $\rv=-0.22$, and $\rv=-0.23$. (Fits for $\rv=-0.21$, and $\rv=-0.23$ are shown in Figs.~\ref{fig:fssFits} and~\ref{fig:fssFitsTc}.)}
    \begin{ruledtabular}
    \begin{tabular}{ccm{2cm}m{2cm}m{2cm}m{2cm}m{2cm}m{2cm}d}
        \textrm{Exponent} & \textrm{Eq./}& \multicolumn{2}{c}{\textrm{$\rv = -0.21$}}& \multicolumn{2}{c}{\textrm{$\rv = -0.22$}} & \multicolumn{2}{c}{\textrm{$\rv = -0.23$}} & \multicolumn{1}{c}{\textrm{Onsager}} \\
        &\textrm{Ref.}& \multicolumn{1}{l}{\textrm{Exponent}}& \multicolumn{1}{l}{\textrm{$\beta_c$ from Eq.~\eqref{eq:TcFitAnsatz}}} & \multicolumn{1}{l}{\textrm{Exponent}}& \multicolumn{1}{l}{\textrm{$\beta_c$ from Eq.~\eqref{eq:TcFitAnsatz}}} & \multicolumn{1}{l}{\textrm{Exponent}}& \multicolumn{1}{l}{\textrm{$\beta_c$ from Eq.~\eqref{eq:TcFitAnsatz}}} &  \\ \colrule
        $\gamma / \nu$ &Eq.~\eqref{eq:fss_chi_max}& 1.728(7) &3.2542(6) & 1.722(7) &4.2693(8)& 1.720(7) &6.3436(11)& 1.75 \\
        &Eq.~\eqref{eq:fss_chi_betaC}& 1.769(6)$^\dagger$ && 1.759(6)$^\dagger$ && 1.734(14)$^\dagger$ && \\
        &Ref.~\cite{Gessert2024}& 1.764(9)$^{\dagger,\ddagger}$ && 1.756(4)$^{\dagger,\ddagger}$ && --- && \\
        $1 / \nu$&Eq.~\eqref{eq:fss_dlm} & 0.96(1) &3.2561(13)& 0.98(1) &4.2703(18)& 0.98(1) &6.346(3)& 1.0 \\
        &Eq.~\eqref{eq:fss_dU4db}&0.91(4)&3.254(3)&0.93(4)&4.265(3)&0.96(3)&6.345(4)&\\
        &Ref.~\cite{Gessert2024}&  0.961(6)$^{\ddagger}$ &3.2572(5)&  0.970(5)$^{\ddagger}$ &4.2702(6)& --- &---& \\
        $\frac{1 - \beta}{\nu}$ &Eq.~\eqref{eq:fss_dm}& 0.869(9) &3.2570(8)& 0.879(9) &4.2712(10)& 0.862(9) &6.3455(14)& 0.875\\ 
        $\alpha$ &Eq.~\eqref{eq:fss_CV}& $=0$ &3.2591(13)& $=0$ &4.2729(14)& $=0$ &6.349(3)& 0.0 \\
        $\beta / \nu$ &Eq.~\eqref{eq:fss_m_max}& 0.117(13) && 0.115(12)&& 0.115(13) && 0.125 \\
        &Eq.~\eqref{eq:fss_m_betaC}& 0.114(4)$^\dagger$ && 0.119(4)$^\dagger$ && 0.135(9)$^\dagger$ &&
    \end{tabular}
    \begin{flushleft}
      $^\dagger$ denotes results from fits at $\beta_c$. Otherwise, fits are at pseudocritical points. For $\rv=-0.21$ and $-0.22$ our estimates for $\beta_c$ of Ref.~\cite{Gessert2024} are used, and for $\rv=-0.23$ the estimate from the scaling of $\beta_{\max}^{(\dlm)}$ is employed.

    \noindent $^\ddagger$ denotes results from Ref.~\cite{Gessert2024} obtained from the scaling of the partition function zeros. $\gamma/\nu$ is calculated from the fit result of the Lee-Yang zeros, and $1/\nu$ from the Fisher zeros. In both cases the fitting range $24$-$88$ was selected such that the degrees of freedom are identical to the fits here.
    \end{flushleft}
\end{ruledtabular}
\label{tab:fssExponents}
\end{table*}
\endgroup
\begin{figure}[ht]
    \includegraphics{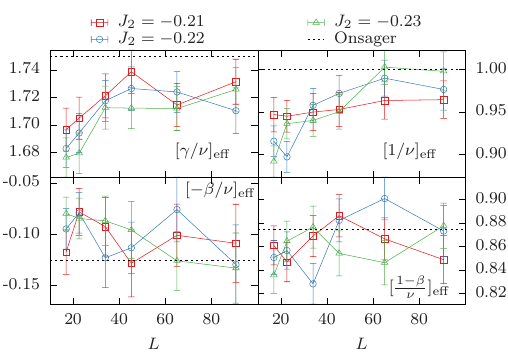}
    \caption{Effective exponents as a function of $L$ for $\rv \in \{-0.21,-0.22,-0.23\}$. Particularly, for the exponents $\gamma/\nu$ and $1/\nu$ a variation with system size is seen, which is indicative of the presence of corrections-to-scaling. The dashed lines denote the asymptotically expected Onsager exponents.\label{fig:fssCorrections}}
\end{figure}

The obtained critical exponents and inverse critical temperatures are summarized in Table~\ref{tab:fssExponents}. Almost all obtained exponents are within one or two standard deviations of the expected Onsager exponents listed in the right column. This suggests that the model remains in the Ising universality class even rather close to the special point $\rv=-1/4$. 
In contrast to the values of $1/\nu$ obtained from the FSS fits of the peak locations of $\dlm$, the ones obtained via $\dU4$ are particularly far from the expected value of 1.
This may be due to $\dU4$ exhibiting broader maxima than $\dlm$ for the same $L$ in the present model (not shown), which also is reflected in much larger error bars for the obtained values of $\beta_{\max}^{\dU4}$ in Fig.~\ref{fig:fssFitsTc}, particularly for smaller system sizes. Also note that the values for $\dU4$ outside the fitting range in Fig.~\ref{fig:fssFits}, i.e., data for $L<48$, fall well above the fitting line, suggesting the presence of strong corrections to scaling, likely still affecting the presented estimates of $1/\nu$ using Eq.~\eqref{eq:fss_dU4db} in Table~\ref{tab:fssExponents}.
Another exception are the $\gamma/\nu$ exponents from the fit for the peaks of the magnetic susceptibility with subtraction.  It is well known that using the magnetic susceptibility with subtraction (especially of the finite-size type $\langle |m|\rangle$) is subject to strong corrections (see, e.g., Refs.~\cite{holm:93a,zierenberg:17,akritidis:22,akritidis:23}). The disconnected form $\chi'$ for $\beta\leq \beta_c$ evaluated at $\beta=\beta_c$ commonly shows better scaling, as is also the case here.
Nonetheless, even the $\gamma/\nu$ values from $\chi_L'(\beta_c$) are about two to three standard deviations away from $1.75$.
For $\rv=-0.21$ and $-0.22$ we use our $\beta_c$ estimates of Ref.~\cite{Gessert2024}, i.e., $\beta_c = 3.2572(5)$ and $4.2702(6)$. For $\rv=-0.23$ we use $6.346(3)$
from the scaling of the location of $\dlm$, i.e., $\beta_{\max}^{(\dlm)}$. Similarly, evaluating the magnetization at $\beta_{\max}^{(\dlm)}$ and at $\beta_c$ gives different estimates for the exponent ratio $\beta/\nu$. Table~\ref{tab:fssExponents} also lists the results from the partition function zeros studied in Ref.~\cite{Gessert2024} for $\rv=-0.21$ and $-0.22$, which, within error bars, are compatible with the values for $\gamma/\nu$, $1/\nu$, and $\beta_c$ obtained here.

These mixed fitting results appear to be due to strong corrections to scaling. One way to examine such scaling corrections is through the effectively varying power-law exponent (corresponding to the local slope in Fig.~\ref{fig:fssFits}), which can be quantified by
\begin{equation}
    [x]_\text{eff} = \frac{\mathrm{d} \ln \mathcal{O}}{\mathrm{d} \ln L},\label{eq:effExpFSS}
\end{equation}
with $x$ and $\mathcal{O}$ being placeholders for a quantity $\mathcal{O}$ with the (asymptotic) FSS relation $\mathcal{O} \propto L^{x}$.
Figure~\ref{fig:fssCorrections} shows the estimated slopes using the three-point midpoint approximation~\footnote{The derivative is approximated using
 \[ [x]_\text{eff}(\tilde{L}_i) = \frac{\ln {O(L_{i+1})} - \ln {O(L_{i-1})}}{\ln L_{i+1}-\ln L_{i-1}},\]
 where $\tilde{L}_i=\exp[(\ln L_{i-1} + \ln L_{i+1})/2]=\sqrt{L_{i-1} L_{i+1}}$ is the geometric mean of $L_{i-1}$ and $L_{i+1}$.} for Eq.~\eqref{eq:effExpFSS}.
Despite the large statistical fluctuation of the numerical derivative, a clear variation of the effective exponents with $L$ can be seen for $\gamma/\nu$ and $1/\nu$ (obtained using the peak locations of $\dlm$). For the largest $L$ the slopes are always consistent with the expected asymptotic exponent. 
Note that the error bars of the effective exponents $[-\beta/\nu]_\text{eff}$ and  $[(1-\beta)/\nu]_\text{eff}$ are too large to precisely evaluate the relevance of scaling corrections.

\subsection{Energy barriers}
\label{sec:energyBarriers}

Finally, we consider the frustration-induced energy barriers present in this model, which form the building blocks of the rough free-energy landscape, and which turn out to provide a rather intuitive explanation of why our rejection-free approach outperforms the standard Metropolis algorithm for low $\rv$, and also why the approach fails for even lower $\rv$. The metastable states reported in Ref.~\cite{Zukovic2021} consist of multiple ferromagnetic domains with extremely long lifetimes. As $\rv$ is reduced the energetic barrier to overturn domains increases as does their lifetime. At the same time the energy difference between the ground state and a state with an overturned domain becomes smaller, rendering it less likely for the excited states to be filtered out during the resampling step in PA.

\begin{figure}
    \includegraphics{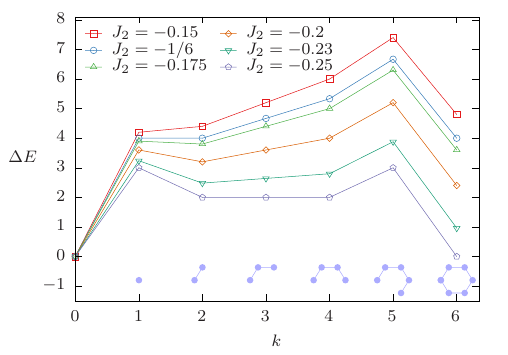}
    \includegraphics{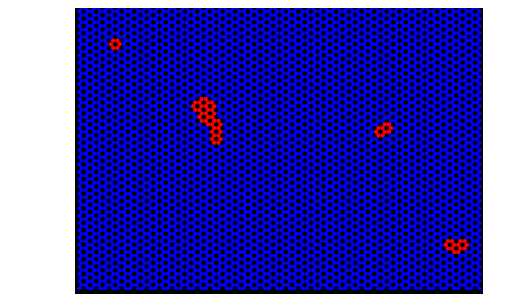}
    \caption{Top: Difference in energy between the excited states and the ferromagnetic ground state for various $\rv$. $k$ denotes the number of overturned spins from the ground state. $k=6$ corresponds to one overturned hexagon (see illustration in the top right). For the calculation of $\Delta E$ we refer to Appendix~\ref{app:brokenBondsInHexExcitation} and Table~\ref{tab:hexExcEnergies} therein. Bottom: Partial snapshot of an $L=64$ system for $\rv=-0.23$ and $\beta\approx 1.2\ \beta_c$.}
    \label{fig:hexExcitation}
\end{figure}

The simplest such domain on the honeycomb lattice corresponds to flipping a single hexagon. In fact, studying snapshots of low-temperature simulations reveals that indeed single hexagonal excitations happen frequently, and that they have a long lifetime; see the bottom panel of Fig.~\ref{fig:hexExcitation}. For the following discussion we consider the difference in energy $\Delta E$ when overturning $k$ consecutive spins on the hexagons (see the top panel of Fig.~\ref{fig:hexExcitation}). Higher excitations, e.g., branching or overturning nonconsecutive spins on the hexagon are ignored, resulting in a very simple one-dimensional representation of the energy landscape.

Starting from the $k=6$ state (cf.\ the top panel of Fig.~\ref{fig:hexExcitation}), the $n$-fold way update brings the system very efficiently to the $k=5$ state, as other available moves are typically even more expensive. Using regular Metropolis, the $6 \rightarrow 5$ move depending on the temperature (which generally is lower the smaller $\rv$) might be rejected many times before eventually being accepted.
Note that the jump in energy from $k = 6$ to $k = 5$ increases with decreasing $\rv$, in line with the general statement about barriers and lifetimes provided above.

From the $k=5$ state, both $\Delta E(5\rightarrow 6)$ and $\Delta E(5\rightarrow 4)$ are negative and thus are always accepted when using Metropolis. As $5\rightarrow 4$ can be realized by flipping a spin on either end of the chain, it is twice as likely to be proposed as $5 \rightarrow 6$. Hence, ignoring any other possible excitations, both with Metropolis and the rejection-free algorithm, the system immediately moves from the state $k=5$ to the $k=4$ state with $2/3$ probability, and back to the $k=6$ state with $1/3$ probability.

Next, from the $k=4$ state, for all $\rv > -0.25$, there is no energetic barrier between $k=4$ and $k=2$. Thus, either method most likely immediately carries out $4\rightarrow 3$ and $3 \rightarrow 2$. What follows depends on $\rv$. For $\rv > -1 / 6$, there are no energy barriers between $k=2$ and the ground state ($k=0$), and both methods likely will go from $k=4$ to $k=0$ in four steps. For $\rv < -1 / 6$, the move $2\rightarrow 1$ increases the energy, making the $k=2$ state a local minimum. As a result, Metropolis simulations may become stuck at $k=2$ for some time, whereas when using $n$-fold way updates one proceeds immediately either to the $k=1$ or the $k=3$ state. In fact, $\rv = -1 / 6$ also experimentally marks the point up to which we easily obtained good simulation data with standard Metropolis. Once in the $k=1$ state, for $\rv < -1 / 6$ both $\Delta E (1 \rightarrow 0)$ and $\Delta E (1 \rightarrow 2)$ are negative and are always accepted. As there are three possible ways to go from $k=1$ to $k=2$ and only one to go to $k=0$, $1 \rightarrow 0$ happens with probability $1/4$ and $1 \rightarrow 2$ with probability $3/4$.

For $\rv < -1 / 5$ the energy difference $\Delta E(2 \rightarrow 1)$ is larger than the one to go from $k=2$ to $k=3$, making the latter more likely to occur for the $n$-fold way update as well as for Metropolis. What happens is that the simulation oscillates between $k=2$ and $k=3$ for some time before reaching $k=1$. The further $\rv$ is reduced, the larger $\Delta E(2 \rightarrow 1)$, and the smaller $\Delta E(2 \rightarrow 3)$ becomes. Hence, as $\rv$ is reduced, reaching $k=1$ takes increasingly long. For example, at $T_c$ for $J_2=-0.23$ going from $k=2$ to $k=3$ is about 45 times more likely than going to $k=1$. Once reaching $k=1$ the chance of selecting $k=0$ next is $1/4$~\footnote{Both $k=0$ and $k=2$ are energetically more   favorable than $k=1$. As there are three nearest neighbors, there are three adjacent states corresponding to $k=2$, and only one state corresponding to $k=0$.}. Thus, the system on average jumps between $k=1$, $2$, and $3$ almost 200 times before reaching the ground state. This is why the $n$-fold way can still equilibrate the system for $J_2=-0.23$, but not as easily as for higher $\rv$. In contrast, for $\rv=-0.24$ assuming $T_c \approx 0.1$~\footnote{Using exact enumeration (see Fig.~\ref{fig:critTemp}) of a system with 32 spins, we find that   the location of the specific-heat peak is at $0.1$. Thus, the actual critical temperature is well below $0.1$ and the $3000$ underestimates the number of spin   flips necessary to reach the $k=1$ state.} we find that the $k=2$ to $k=3$ transition is at least 3000 times more likely than $2 \rightarrow 1$. Thus, ultimately, the $n$-fold way PA cannot equilibrate $\rv=-0.24$ in any reasonable time.
By deriving the transition matrix of this simplified model with seven states and considering its discrete phase-type distribution~\cite{Bladt2017}, one finds that it would take on average 500 spin flips ($2.3\times 10^8$ MCS) to go from $k=6$ to $k=0$ at $T_c$ for $\rv=-0.23$. For $\rv=-0.24$ at $T=0.1$, on the other hand, going from $k=6$ to $k=0$ would require $35\,000$ spin flips ($1.6\times 10^{13}$ MCS).

In addition, it is clear that the closer the energy of the configuration with the fully overturned  hexagon (cf.\ $k=6$ in Fig.~\ref{fig:hexExcitation}) is to that of the ground state, the less likely it becomes that the resampling step in the PA algorithm filters these configurations out.

\section{Summary and Outlook}
\label{sec:conclusion}

We have studied the frustrated $J_1$-$J_2$ Ising model on the honeycomb lattice by utilizing population annealing Monte Carlo simulations. We have carried out a finite-size scaling analysis for $\rv=-0.21$, $-0.22$, and $-0.23$, which in all cases yielded numerical values for the critical exponents consistent with the Ising universality class. Our results are in good agreement with recent studies~\cite{Zukovic2021,Gessert2024} that carried out FSS for larger values of $\rv$. Thus, we conclude that the model remains in the Ising universality class at least for $J_2\geq -0.23$, and very likely also for values below $-0.23$ where our simulations failed to equilibrate. For values of $\rv$ between $-0.23$ and $-0.25$ simulations fail because of energy barriers between the ground state and states with energies slightly above the ground-state energy, but we have not seen anything to signal a change in the type of transition.

In order to equilibrate our simulations we have replaced the standard Metropolis algorithm in the population annealing algorithm by the rejection-free $n$-fold way update~\cite{Bortz1975}, which allowed us to equilibrate lower $\rv$ than in previous work~\cite{Zukovic2021} where low acceptance rates led to quasifrozen states. This shows, on the one hand, that the $n$-fold way update is not just a useful tool in nonequilibrium settings at low temperature but also well suited in equilibrium when no efficient cluster update is available. On the other hand, it highlights that population annealing is agnostic to the used update mechanism which previously was already realized in the context of molecular dynamics simulations~\cite{Christiansen2019a}. This approach is further enhanced by introducing an adaptive number of sweeps into the population annealing framework, which works particularly well for the model at hand: Energy barriers cause simulations to be trapped in local energy minima at low temperatures. By using the adaptive sweep protocol, the population is quickly cooled at high temperatures when equilibration is easy and computing cost is focused on the temperature range where simulations tend to be trapped in these minima.

Our method still fails to equilibrate at even smaller values for $\rv$, viz.\ $\rv \in (-0.25, -0.23)$. We attribute this to the fact that the energies of configurations containing overturned domains are close to the ground-state energy and that these configurations are separated from the ground state by large energetic barriers.
This, on the one hand, makes it difficult for PA's resampling to filter out configurations containing such domains, and, on the other, results in those domains having very long lifetimes. $\rv$ even closer to (but larger than) $-0.25$ could possibly be simulated by taking into account these excitations, for example by proposing to flip two neighboring sites (thus eliminating the $2\leftrightarrow 3$ oscillation in
Fig.~\ref{fig:hexExcitation}) or to flip entire hexagons. (An adaptation of the $n$-fold way algorithm to include updates of multiple spins simultaneously was proposed in Ref.~\cite{Novotny1995}.) %

Both, using the rejection-free update and the adaptive sweep schedule, as well as the combination thereof may be useful for studying other systems in which rugged free-energy landscapes and low acceptance rates render numerical simulations difficult. %
However, for our method to be suitable as a general-purpose approach, two issues require addressing. First, advancing all replicas to the maximum of the  observed Metropolis times among them can deteriorate parallel efficiency when all replicas have to wait for one or a few other replicas. As the maximum is used, this problem becomes worse the larger the population size is. While this issue did affect our simulations to some extent, the extreme-value distribution of the Metropolis times in our case still allowed for efficient computation for the chosen population size. Second, the adaptive sweep protocol suggested here assumes that it is possible to decorrelate replicas through Monte Carlo updates. Clearly, when ergodicity is effectively broken, as is typically the case in glassy systems below the glass-transition temperature, the decorrelation criterion may not be satisfied within a reasonable time, and the simulation likely would fail to proceed~\footnote{In fact, this is exactly the fate of our $\rv=-0.24$ simulations.}. In such cases the recently proposed adaptive sweep schedule~\cite{Barzegar2024} using the ``restricted autocorrelation time'' may be more suitable. %

\emph{Note added.} After the submission of this paper we became aware of a related Wang-Landau simulation study by Azhari~\textit{et al.}~\cite{Azhari2025}, which also investigates the critical behavior of this model. %

\begin{acknowledgments}
    This project was supported by the Deutsch-Franz\"osische Hochschule (DFH-UFA) through the Doctoral College ``$\mathbb{L}^4$'' under Grant No.\ CDFA-02-07. We further acknowledge support by the Leipzig Graduate School of Natural Sciences ``BuildMoNa''. M.W. acknowledges support from Emory University for a sabbatical stay, where part of this work was completed.

    Most calculations were performed using the Sulis Tier 2 HPC platform hosted by the Scientific Computing Research Technology Platform at the University of Warwick. Sulis is funded by EPSRC Grant No.\ EP/T022108/1 and the HPC Midlands+ consortium.
\end{acknowledgments}

\appendix

\vspace*{1em}

\begin{figure*}[ht]
    \includegraphics[width=\textwidth]{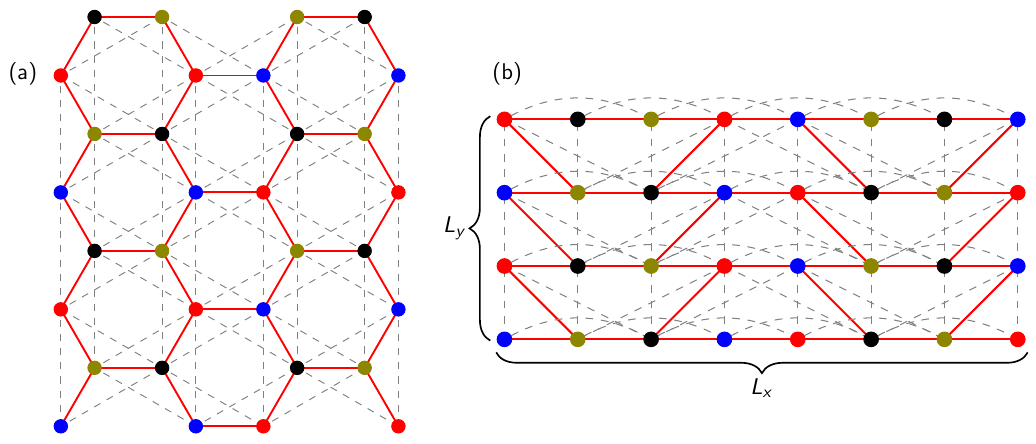}
    \caption{Domain decomposition into four subsets of noninteracting spin sites for the $J_1$-$J_2$ Ising model on the honeycomb lattice of linear system size $L=4$ containing $N=2L^2=32$ sites. Each color refers to a different subset. (a) Representation in ``real'' coordinates, and (b) in ``memory'' coordinates, showing how spins are arranged in a two-dimensional array structure $L_x\times L_y$ with $L_x=2L$ and $L_y=L$.\label{fig:honeycombDomainDecomposition}}
\end{figure*}
\section{GPU domain decomposition}\label{app:domainDecompositionHoneycomb}
Since the sites are arranged regularly in the considered honeycomb model, and since each spin only interacts with a small number of spins within its vicinity, a domain decomposition in the spirit of the well-known checkerboard decomposition~\cite{Heermann1990} of the square lattice with even side lengths $L$ is possible. Due to the {\nn} and {\nnn} interactions, it is not possible to decompose the lattice into two sublattices of noninteracting sites. Instead, four colors are required, but otherwise the procedure is analogous to the one of the checkerboard decomposition; see Fig.~\ref{fig:honeycombDomainDecomposition}(a). In memory, the spins are represented in a two-dimensional array: In Fig.~\ref{fig:honeycombDomainDecomposition}(b) the same decomposition is displayed in ``memory'' coordinates. Using this decomposition, one sweep consists of updating i) all blue sites, ii) all red sites, iii) all olive sites, and iv) all black sites (or any other order thereof), where in each step $N/4$ spins are updated in parallel.

\section{Binder parameter crossings}\label{app:binderCrossing}
\begin{figure}
    \includegraphics{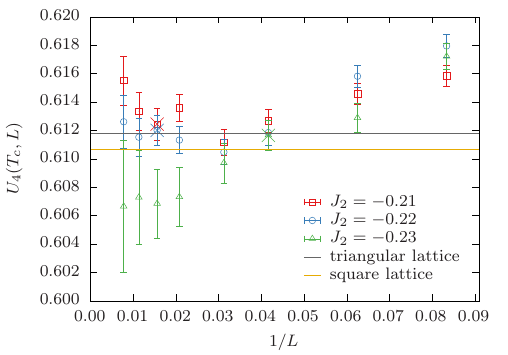}
    \includegraphics{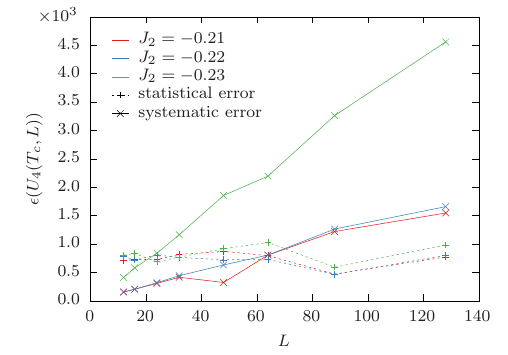}
\caption{Binder parameter $U_4$ at $T_c$ for different $\rv$ and $L$. Top: $U_4$ vs $1/L$ at $T_c$. The error bars correspond to the square root of the sum of the squared systematic and statistical errors. The values quoted in the text are highlighted by large crosses ($\times$). 
As reference values the exact values~\cite{Kamieniarz1993} of the triangular and the square lattice are indicated by colored lines. Bottom: The statistical error (dashed lines) and the systematic error (solid lines) as a function of $L$ for different $\rv$ (indicated by different colors).\label{fig:errBinderCross}}
\end{figure}

In contrast to the critical exponents, $U^*$ is only weakly universal, i.e., it depends on details of the lattice structure. 
The numerical estimates for $U^*$ quoted in the main text were produced as follows. $U_4(T_c,L)$ was determined for the different system sizes $L$ at the previously obtained values of $T_c$ using (single) histogram reweighting from the closest simulation point; see the top panel of Fig.~\ref{fig:errBinderCross}. Besides statistical fluctuations in $U_4$ and (unknown) finite-size corrections affecting the estimates for $U^*$, the inaccuracy of $T_c$ results in an additional systematic error contribution.
Assuming that the two contributions are independent, the displayed error bars were obtained as the square root of the sum of the squared error contributions, that is, for example, the error bar of $0.0012$ for $\rv=-0.21$ and $L=64$ contains a contribution from the statistical error of $0.00080$ and one from the systematic error of $0.00081$.
As can be seen in the top panel of Fig.~\ref{fig:errBinderCross}, for small~$L$ the $U_4$ values differ significantly from the rest due to finite-size corrections, whereas for larger systems the estimate is affected by the increasing total error.
Thus, the system size at which $U^*$ is determined has to be selected somewhat carefully.

Due to how the simulations are set up, the statistical error is roughly system-size independent, whereas the systematic error scales with the slope of $U_4$ at $T_c$, i.e., with $L^{1/\nu}=L$, which is the reason for the error bars increasing with the system size in the top panel. 
This effect can be seen in the bottom panel of Fig.~\ref{fig:errBinderCross}. The quoted values for $U^*$ in the main text, i.e., $0.6124(12)$, $0.6120(11)$, and $0.6117(11)$ for $\rv=-0.21$, $-0.22$, and $-0.23$, were obtained using $U_4(T_c,L)$ of the smallest $L$ for which the systematic error due to the uncertainty in $T_c$ is greater than the statistical error. (The data points used for the final estimates for $U^*$ are highlighted in the top panel by large crosses.) For $\rv=-0.21$ and $-0.22$ this $L$ is 64, and for $\rv=-0.23$ it is 24 (the difference being due to the different accuracy in $T_c$). 
Note, that these values are within error bars of the value of the pure Ising model on the triangular lattice, $U^*=0.611827739(14)$ (Ref.\ \cite{Kamieniarz1993}), but in any case even the (different) value $U^* = 0.6106901(14)$ of the square lattice~\cite{Kamieniarz1993} is rather close.

~

\section{Energies for six states leading to the hexagonal excitation}\label{app:brokenBondsInHexExcitation}
The energies discussed in Sec.~\ref{sec:energyBarriers} can be calculated in a straightforward manner by counting the number of broken nearest- and the number of satisfied next-nearest-neighbor bonds for the six states $k=1,\dots,6$. These numbers are compiled in Table~\ref{tab:hexExcEnergies}.

\begin{table}[ht]
    \caption{Number of broken FM nearest-neighbor bonds and satisfied AF next-nearest-neighbor bonds for the states leading to the hexagonal excitation. The difference in energy of the state $k$ and the ferromagnetic ground state is given by $\Delta E(k) = 2 \times [N_1(k) J_1 + N_2(k) J_2]$.}
    \begin{ruledtabular}
    \begin{tabular}{cdd}
        $k$ & \multicolumn{1}{c}{\textrm{$N_1 = \# J_1$} bonds broken} & \multicolumn{1}{c}{\textrm{$N_2 = \# J_2$} bonds satisfied} \\ \colrule
        $0$ & 0 & 0\\
        $1$ & 3 & 6 \\
        $2$ & 4 & 12 \\
        $3$ & 5 & 16 \\
        $4$ & 6 & 20 \\
        $5$ & 7 & 22 \\
        $6$ & 6 & 24
    \end{tabular}
\end{ruledtabular}
\label{tab:hexExcEnergies}
\end{table}

\bibliography{bibliography}

\end{document}